\documentclass[12pt, secnumarabic, amssymb, nofootinbib, aps, prd]{revtex4}

\usepackage[english]{babel}
\usepackage{amsmath}
\usepackage{amssymb}
\usepackage{amsbsy}
\usepackage{amstext}
\usepackage{graphicx}
\usepackage{pst-node}
\usepackage{verbatim}
\usepackage{tikz}
\usetikzlibrary{decorations.pathmorphing}
\usetikzlibrary{shapes,arrows,positioning,automata,backgrounds,calc,er,patterns}
\usepackage{tikz-feynman}
\usepackage{ytableau}
\makeatletter\AtBeginDocument{\let\@elt\relax}\makeatother

\newcommand{\be}{\begin{eqnarray}}
\newcommand{\ee}{\end{eqnarray}}
\newcommand{\bdm}{\begin{displaymath}}
\newcommand{\edm}{\end{displaymath}}
\newcommand{\ds}{\displaystyle}
\newcommand{\nn}{\nonumber}
\newcommand{\ta}{\tt{a}}
\newcommand{\ba}{\begin{array}}
\newcommand{\ea}{\end{array}}
\newcommand{\pa}[1]{\left(#1\right)}
\newcommand{\paq}[1]{\left[#1\right]}
\newcommand{\pag}[1]{\left\{#1\right\}}

\newcommand{\K}{{\bf k}}
\newcommand{\Q}{{\bf q}}
\newcommand{\pp}{{\bf p}}
\newcommand{\X}{{\bf x}}

\newcommand{\R}{{\bf r}}

\begin{document}

\title{Tail contributions to gravitational conservative dynamics}

\author{Gabriel Luz Almeida}
\email{gabriel.luz@fisica.ufrn.br}
\affiliation{Departamento de F\'\i sica Te\'orica e Experimental, Universidade Federal do Rio Grande do Norte, Avenida Senador Salgado Filho, Natal-RN 59078-970, Brazil}

\author{Stefano Foffa}
\email{stefano.foffa@unige.ch}
\affiliation{D\'epartement de Physique Th\'eorique and Center for Astroparticle Physics, Universit\'e de Gen\`eve, CH-1211 Geneva, Switzerland}

\author{Riccardo Sturani}
\email{riccardo.sturani@ufrn.br}
\affiliation{International Institute of Physics, Universidade Federal do Rio Grande do Norte, Campus Universit\'ario, Lagoa Nova, Natal-RN 59078-970, Brazil}

\begin{abstract}
  We compute \emph{tail} contributions to the conservative
  dynamics of a generic self-gravitating system, for every
   multipole order, of either electric and
  magnetic parity.
  Such contributions arise when gravitational radiation
  is backscattered by the static curvature generated by the source itself and
  reabsorbed.
  Tail processes are relevant for the study of compact binary dynamics 
    starting from the fourth post-Newtonian order.
 Within this context, we resolve a discrepancy recently arised at fifth
  post-Newtonian order between effective field theory and self-force results.
\end{abstract}

\maketitle


\section{Introduction}

The recent detections of gravitational waves from coalescing binaries
\cite{LIGOScientific:2018mvr,LIGOScientific:2020ibl,LIGOScientific:2021qlt}
by the large interferometers LIGO \cite{TheLIGOScientific:2014jea} and Virgo
\cite{TheVirgo:2014hva} have not only opened a new field in astronomy, but
also revived theoretical studies of the general relativistic two-body problem.

Starting from the multipole expansion of a generic source, the focus of the
present work is the contribution to the conservative dynamics of certain types
of nonlinear effects, called \emph{tails}, which are quadratic in Newton constant $G_N$
and arise when radiation is emitted, scattered onto the static potential
generated by the same source, and finally absorbed.
Tail effects have been first identified in emission processes, i.e.,
in the scattering of radiation off the source background curvature
\cite{Blanchet:1987wq},
and the leading contribution to the conservative dynamics of compact binaries has been first quantified
in \cite{Foffa:2011np}; see also \cite{Galley:2015kus}.

A common framework for dealing with two-body dynamics is the post-Newtonian
(PN) approximation to General Relativity (GR) (see \cite{Blanchet:2013haa} for
a review)
that treats perturbatively the two-body problem and whose parameter of
expansion is the binary relative velocity $v$, where the third Kepler's law
ensures $v^2\sim G_NM/r$, where $M$ is the binary system total mass and $r$ is the
binary constituents' distance.
The PN approximation scheme is almost as old as GR itself; however, remarkably,
it has been recast in recent times by \cite{Goldberger:2004jt} in a purely effective field theory (EFT)
framework, dubbed nonrelativistic GR (NRGR) as it borrowed ideas from
nonrelativistic QCD.

A technical detail of the NRGR framework, which is a common tool in field theory
computations, is the distinction of the momenta of modes exchanged between
sources, i.e., gravitational modes in NRGR, according to the \textit{region} they belong to: \textit{hard} or \textit{soft}.

\emph{Potential} modes sourced by conserved charges are of the former type,
which are characterized by 4-momenta $k\sim (k_0,\K)$ with $|\K|\sim 1/r\gg k_0\sim v/r$,
while \emph{radiative} modes sourced by radiative multipoles have $|\K|\sim k_0$
and belong to the soft region.
This separation is related to the distinction in a \emph{near} and a
\emph{far} zone, commonly adopted in the \emph{traditional} PN approach,
where in the former binary constituents are resolved
as individual objects and in the latter the binary system is described
as a composite source endowed with multipoles.
The complete dynamics is given by the sum of near and far zone processes
\cite{Foffa:2021pkg}, as it is well known from the particle physics application
of the method of regions \cite{Manohar:2006nz,Jantzen:2011nz}.

The tail contributions presented in this work are linear in $M$ and quadratic
in the nonconserved multipoles. The tail process involving electric quadrupole
moments concur to determine the
4PN conservative dynamics of two-body system, which has now been obtained with
three different methods independently by several groups \cite{Bini:2013zaa,Damour:2014jta,Bernard:2015njp,Marchand:2017pir,Foffa:2019rdf,Foffa:2019yfl,Blumlein:2020pog}
with mutually agreeing results.

Note that for obtaining the correct two-body dynamics, it is not enough to
compute the tail effects in terms of multipoles, but also to
derive the appropriate expression of the multipoles in terms of individual
constituents, in a procedure that in the EFT language is called \emph{matching}.

Since tail processes involve regularization of divergences, which we treated in
dimensional regularization, to obtain the correct result it is crucial to have a
multipole expression which is correct in generic $d$ dimensions.
While this is straightforward for electric multipoles \cite{Blanchet:2005tk}, it involves some
subtleties for magnetic ones, which are often expressed in terms of the
Levi-Civita antisymmetric tensor $\epsilon_{ijk}$, which is not
straightforwardly generalizable to $d\neq 3$.
Besides computing in the present work for the first time tail terms for all
multipoles, we solve the puzzle arised by the apparent discrepancy between the
conservative dynamics NRGR result at 5PN, obtained in \cite{Foffa:2019eeb} for the far zone and in \cite{Blumlein:2020pyo}
for the potential modes, and the derivation of the two-body scattering angle from the ``Tutti Frutti'' approach
\cite{Bini:2019nra,Bini:2020hmy}, as presented in \cite{Bini:2021gat}, at next-to-leading
order in symmetric mass ratio $\nu=m_1m_2/M^2$,
having denoted by $m_{1,2}$ the masses of the binary constituents.
We show here that agreement is indeed found between the two approaches when
magnetic multipoles are properly defined in generic $d$ dimensions, as in \cite{Henry:2021cek}.

The outline of the paper is as follows: in Sec.~\ref{sec:calculation}
we perform the computation of the tail process for generic multipoles,
completing it in Sec.~\ref{sec:matching} with the necessary 
matching procedure, which enables one to write the multipoles in terms of the
binary constituents' variables. In Sec.~\ref{sec:comparison} we explain the origin
of the discrepancy found in \cite{Blumlein:2020pyo,Bini:2021gat} and show its
solution, and Sec.~\ref{sec:conclusion} contains the conclusions that can be
drawn from our results.

\section{Far zone calculations}
\label{sec:calculation}

\begin{figure}
  \begin{center}
  \includegraphics[width=.5\linewidth]{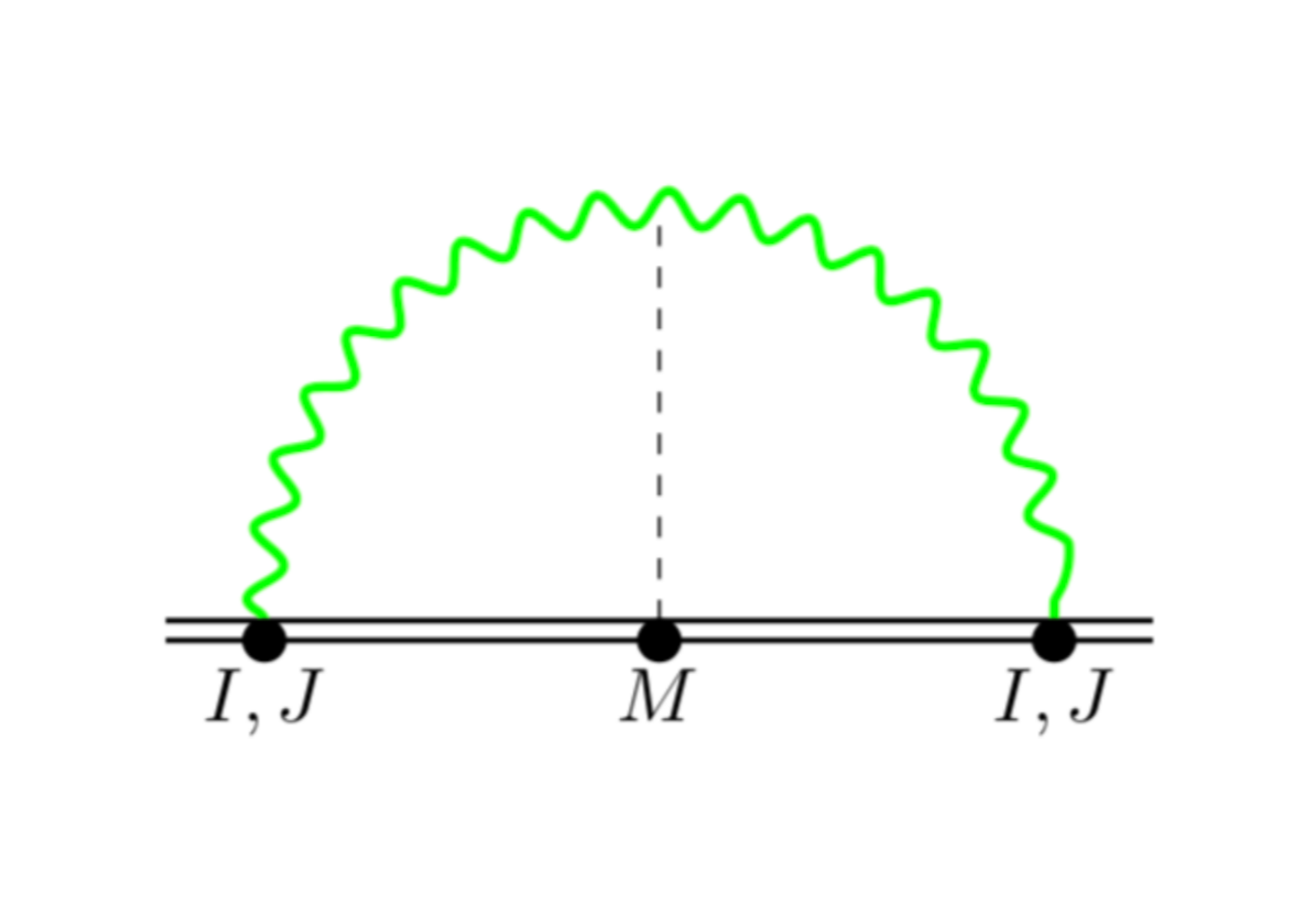}
  \end{center}
  \caption{Self-energy diagram for a generic radiative multipole source of electric and magnetic type.
    The green wavy (dashed straight) line represents the radiative (potential)
    gravitational field, while the black double line stands for the composite source
    described by the multipolar action in Eq.~(\ref{eq:smult}).}
  \label{fig:self_energy_tail}
\end{figure}

We adopt the following parametrization for the multipolar expansion of the gravitational coupling of a composite system:
\begin{align}
 \label{eq:smult}
{\cal S}_{\text{mult}} &=\int {\rm d}t\left[\frac 12 Eh_{00}-\frac 12 J^{b|a}h_{0b,a} 
     -\sum_{r\geq 0}\left(c^{(I)}_r I^{ijR}\partial_RR_{0i0j}+\frac{c^{(J)}_r}2 J^{b|iRa}\partial_RR_{0iab}\right)
     \right]\,,
\end{align}

with \cite{Thorne:1980ru,Ross:2012fc}
\be
c_r^{(I)}=\ds\frac 1{\pa{r+2}!}\,,\quad c_r^{(J)}=\ds\frac {2\pa{r+2}}{\pa{r+3}!}
\,,
\ee
where $E$ denotes the energy of the system and $h_{\mu\nu}$ indicates the metric
deviation from Minkowski, i.e., $g_{\mu\nu}=\eta_{\mu\nu}+h_{\mu\nu}$.\footnote{
  Latin indices $a,\ldots, n$ indicate purely spatial coordinates and we adopt
  the mostly plus metric, $\eta_{\mu\nu}={\rm diag}(-,+,\dots,+)$. Because
  space indices are raised and lowered with a Kronecker delta, we will identify
  covariant and contravariant space indices.}
We denote the generic electric traceless multipoles by
$I^{ijR}\equiv I^{iji_1\dots i_r}$, using the collective index $R$, and the magnetic
traceless multipoles by $J^{b|iRa}\equiv J^{b|ii_1\dots i_ra}$
(antisymmetric under $a\leftrightarrow b$), using the $d$-dimensional
generalization of the magnetic moments recently proposed in
\cite{Henry:2021cek}, which are described in Appendix \ref{app:Jd}.
The tail process we are interested in is described by the diagram in
Fig.~\ref{fig:self_energy_tail}. It represents the leading far zone contribution
to the conservative dynamics, i.e., the leading contribution from diagrams
involving radiative gravitational fields.

The solid double line represents the nonpropagating source of Eq.~\eqref{eq:smult},
characterized by conserved energy and angular momentum, and time-dependent
electric and magnetic multipoles; wavy lines represent radiative mode
Green's function and the dashed line stands for the potential mode Green's
function, meeting at the triple bulk vertex where the scattering of radiation
off the potential happens.
Using standard Feynman rules for the worldline and bulk vertices involved
(see, e.g., \cite{Foffa:2019eeb}) and adopting the notation $p_{R}\equiv p_{i_1\dots i_r}$
and $\int_{\pp\,\Q}\equiv \int\frac{{\rm d}^dp}{(2\pi)^d}\frac{{\rm d}^dq}{(2\pi)^d}$, one has
\be
\label{eq:QQQret}
\ds{\cal S}^{(e,r)}_{tail}&=&\ds-\frac{[c_r^{(I)}]^2 E}{4\Lambda^4}
\int\frac{{\rm d}p_0}{2\pi}\int_{\pp,\Q}\frac{p_{R}q_{R'}
  I^{ijR}(p_0)I^{klR'}(-p_0)}{\pa{\pp^2-p_0^2}\pa{\Q^2-p_0^2}\pa{\Q-\pp}^2}\nn\\
&\times&\left\{-\frac14 p_0^6\delta_{ik}\delta_{jl}-p_0^4\delta_{ik}p_j\pa{q-p}_l-\frac{p_0^2}{c_d}p_i p_j\paq{\pa{q-p}_k\pa{q-p}_l+\frac12 q_k q_l}\right.\nn\\
&&\quad\left.+\frac{p_0^2}2p_i q_k\paq{\pp\cdot\Q\delta_{jl}+p_jq_l-p_lq_j}\right\}\\\
&=&-\frac E{(32 \pi \Lambda^2)^2} \frac{(d-2)\Gamma\pa{d-3}\Gamma\pa{2-\frac d2}\Gamma\pa{-\frac d2 -1-r}}{2^r(4\pi)^{d-2}(d-1)(r+1)^2(r+2)^2\Gamma\pa{d+r}}\zeta^{(e)}_{d,r}\nn\\
&&\ds\quad\times\int\frac{{\rm d}p_0}{2\pi}(-p_0^2)^{d+r}I^{ijR}(p_0)I^{ijR}(-p_0)\nn
\ee
with
\be
\zeta^{(e)}_{d,r}&\equiv& d^4+2(r-2)d^3+ \pa{3 r^3 -3 r +5}d^2+ \pa{r-2}\pa{2 r^2 +r+1}d+\pa{r+1}^2\pa{r^2 + 2 r + 12}\nn\\
&=&(r+1)(r+2)(r+3)(r+4)+(r+4)(2r^2+7 r+7)\varepsilon+{\cal O}\pa{\varepsilon^2}\,,\quad\varepsilon\equiv d-3\,,
\ee
which can be obtained using the master integral reported in Appendix \ref{app:intgrs}; see also \cite{Almeida:2021jyt}.
It has been necessary to introduce a $d$-dimensional Newton constant in the form
$G_d=\mu^{d-3}\pa{32\pi\Lambda^2}^{-1}$, with $\Lambda\equiv (32\pi G_N)^{-1/2}$,
where $\mu$ is an arbitrary, inverse length scale needed to adjust dimensions.

Expansion around $d\sim 3$ gives
\be
\label{eq:e_tail}
\ds{\cal S}^{(e,r)}_{tail}&\simeq& -G_N^2 E\frac{2^{r+2}(r+3)(r+4)}{(r+1)(r+2)(2r+5)!}\int\frac{{\rm d}p_0}{2\pi}(p_0^2)^{r+3}I^{ijR}(p_0)I^{ijR}(-p_0)\paq{\frac1{\tilde{\varepsilon}}-\gamma^{(e)}_r}
\ee
with
\be
\label{eq:eps_def}
\frac1{\tilde{\varepsilon}}\equiv \frac1\varepsilon- {\tt i}\pi + \log\pa{\frac{p_0^2 {\rm e}^\gamma}{\pi \mu^2}}
\ee
and
\be
\ba{rcl}
\label{eq:gammae_def}
\ds\gamma^{(e)}_r&\equiv&\ds \frac12\pa{H_{r+\frac52}-H_{\frac12} +2H_r +1} +\frac2{(r+2)(r+3)}\\
&=&\ds\left\{\frac{41}{30},\frac{82}{35},\frac{1819}{630},\frac{11359}{3465},\frac{646141}{180180},\dots\right\}\,,
\ea
\ee
where $H_r$ is the $r$th harmonic number defined by $H_r\equiv \sum_{i=0}^r 1/i$.
The structure contained in Eq.~(\ref{eq:eps_def}), whose precoefficient in the tail action is the opposite of the one multiplying the multipole contribution to the
flux \cite{Thorne:1980ru} (see below), is universal for all multipoles and has been analyzed in \cite{Foffa:2019eeb}. It has been obtained using Feyman Green's
functions, which give for the real part the same result as it would be obtained by adopting causal boundary conditions, that is, an UV pole, to be matched by an IR pole coming from the near zone, and
the nonlocal (in time) contribution discovered in \cite{Blanchet:1987wq}.
The imaginary part, which is of no interest for the present work, is related
via the optical theorem to the energy flux in tail process; see
\cite{Foffa:2021pkg} for a thorough discussion on the appropriate choice of the boundary conditions.

Coming to the first new result of this work, that is, the $\gamma^{(e)}_r$ term displayed Eq.~(\ref{eq:gammae_def}), we notice that it is rational because the irrational terms contained in
$H_{r+\frac52}$ and $H_{\frac12}$ cancel in the difference $\ds H_{r+\frac52}-H_{\frac12}=2\sum_{i=1}^{r+2}\frac1{2i+1}$.
The first two values $\gamma^{(e)}_0$ and $\gamma^{(e)}_1$  correspond to the electric
quadrupole and octupole finite and local-in-time contributions computed, respectively, in
\cite{Foffa:2011np} and \cite{Foffa:2019eeb}.

An analogous computation for the magnetic moments gives
\be
\ds{\cal S}^{(m,r)}_{\textrm{tail}}
&=&\ds\frac{[c_r^{(J)}]^2 E}{32\Lambda^4} \int\frac{{\rm d}p_0}{2\pi} J_{l|jRk}(p_0) J_{n|aR'm}(-p_0)
\int_{\pp,\Q}\frac{p_{R}q_{R'}}{\pa{\pp^2-p_0^2}\pa{\Q^2-p_0^2}\pa{\Q-\pp}^2}\nn\\
&&\times\ds\bigg\{ p_0^4 p_l q_n \pa{\delta_{ja}\delta_{km}+\delta_{jm}\delta_{ka}}+2p_0^2 p_l q_m p_j \paq{ \delta_{ka}p_n+ \delta_{kn}(p-q)_a}\nn\\
&&\ds\quad- q_a q_m p_j p_k\pa{\pp\cdot\Q\delta_{ln}-q_l p_n} \bigg\}\\
&=&-\frac E{(32 \pi \Lambda^2)^2} \frac{\Gamma\pa{d-3}\Gamma\pa{2-\frac d2}\Gamma\pa{-\frac d2 -1-r}}{2^{r-1}(4\pi)^{d-2}(r+1)(r+3)^2\Gamma\pa{d+r+1}}
\paq{d^2+d(r-2)+r^2+2r+3}\nn\\
&&\quad\times\pa{d+r+1}\int\frac{{\rm d}p_0}{2\pi}(-p_0^2)^{d+r}J_{l|jRk}(p_0) J_{n|aRk}(-p_0) \paq{\delta_{ja}\delta_{ln}+ (r+1)\delta_{jn}\delta_{la}}\,.\nn
\ee
Expansion around $d\sim3$ gives
\be
\label{eq:tailmr}
\ba{rcl}
\ds{\cal S}^{(m,r)}_{tail}&\simeq&\ds -G_N^2 E\frac{2^{r+4}(r+2)(r+4)}{(r+1)(r+3)^2(2r+5)!}\\
&&\ds\quad\times\int\frac{{\rm d}p_0}{2\pi}(p_0^2)^{r+3} J_{l|jRk}(p_0) J_{n|aRk}(-p_0) \paq{\delta_{ja}\delta_{ln}+ (r+1)\delta_{jn}\delta_{la}}\paq{\frac1{\tilde{\varepsilon}}-\gamma^{(m)}_r}
\ea
\ee
with
\be
\ba{rcl}
\ds\gamma^{(m)}_r&\equiv&\ds \frac12\pa{H_{r+\frac52}-H_{\frac12} +2H_{r+3} +2}- \frac{2r^2+13 r+22}{(r+2)(r+3)(r+4)}\\
&=&\ds\left\{\frac{49}{20},\frac{22}{7},\frac{4541}{1260},\frac{1565}{396},\frac{1525229}{360360}\dots\right\}\,.
\ea
\ee
Also in the magnetic case, the numerical coefficient in front of the $\varepsilon$
pole reproduces, for $d=3$, the magnetic emission coefficients which, however,
in the commonly used formula for the energy flux $F$ appears as (see Appendix
\ref{app:intgrs})
\be
\label{eq:flux}
F=G_N\sum_{r\geq 0} \bigg[ f_I(r)\Big\langle\pa{\frac{{\rm d}^{r+3}}{{\rm d}t^{r+3}}I_{ijR}}^2\Big\rangle+
 f_J(r)\Big\langle\pa{\frac{{\rm d}^{r+3}}{{\rm d}t^{r+3}}J_{ijR}}^2\Big\rangle \bigg] \,,
\ee
i.e., they multiply the square of the magnetic multipole $J_{ijR}$, which is
dual to the $J_{l|jRk}$ used above \cite{Henry:2021cek}:
\be
\label{eq:Jmdual}
J_{l|jRk}\stackrel{d=3}{=}\epsilon^{ilk} J_{ijR}\,.
\ee
When substituting in the result of Eq.~(\ref{eq:tailmr}) the relation (\ref{eq:Jmdual}) one finds
$J_{l|jRk} J_{n|aRk} \paq{\delta_{ja}\delta_{ln}+ (r+1)\delta_{jn}\delta_{la}}\stackrel{d=3}{=}(r+3) J_{ijR}^2$ which reproduces the magnetic emission coefficients
first derived by \cite{Thorne:1980ru}. As for the finite part proportional to $\gamma^{(m)}_r$ in
Eq.~(\ref{eq:tailmr}), care is needed to match to the binary constituents'
variables, which will be done in Sec.~\ref{sec:matching}, and to extract the
correct $d\to 3$ limit, which will be done in Sec.~\ref{sec:comparison}.

\section{Matching}
\label{sec:matching}

Following standard GR textbook derivation (see Appendix \ref{app:Jd} for
details) the linear gravitational coupling of an extended source can be
written in terms of the nonconserved source multipoles. Working in the
transverse-traceless (TT) gauge for simplicity, and focusing for the moment
on the electric multipoles, one can write
\be
   {\cal S}_{mult-e}=\frac 12\int{\rm d}t\pa{\frac 12\ddot I_{ij}h_{ij}+
     \frac 16 \ddot I^{ijk} h_{ij,k}+\ldots}\,,
   \ee
   where at lowest order
   \be
   I^{ij}=\int_V T^{00}x^ix^j\,,\\
   I^{ijk}=\int_V T^{00}x^ix^jx^k\,,
   \ee
   where $T^{\mu\nu}$ is the energy-momentum tensor of a generic source and we
   adopted the notation $\int_V\equiv \int {\rm d}^dx$.
   For the magnetic ones, one needs some extra care, in view of the $d$-dimensional
   generalization which is needed for Eq.~(\ref{eq:tailmr}).
   Taylor expanding the standard, generic gravitational coupling $\frac 12 T^{\mu\nu}h_{\mu\nu}$
   at next-to-leading order for the spatial components $T^{ij}$, one obtains (see
   Appendix \ref{app:Jd} for details)
   \be
   \label{eq:tijxk}
   \ba{rcl}
   \ds \frac 12\int_V T^{ij}x^k&=&\ds\frac 16\int_V\pa{T^{ij}x^k+T^{ki}x^j+T^{jk}x^i}+\frac 16\int_V \pa{2T^{ij}x^k-T^{ik}x^j-T^{jk}x^i}\,,\\
&=&\ds\frac 1{12}\frac{{\rm d}^2}{{\rm d}t^2}\pa{\int_V T^{00}x^ix^jx^k}
+\frac 16\frac{\rm d}{{\rm d}t}\paq{\int_V \pa{T^{0i}x^kx^j+T^{0j}x^kx^i-2T^{0k}x^ix^j}}\,.
\ea
\ee
The first term in the second line of Eq.~(\ref{eq:tijxk}) is the electric
octupole contribution, whereas the second term is the magnetic quadrupole one, which is
often written in terms of
\be
\label{eq:magq}
J^{ij}&\stackrel{d=3}{\equiv}&\ds\frac 12\int_V x^mT^{0n}\pa{x^i\epsilon^{jmn}+x^j\epsilon^{imn}}
\,,
\ee
which is symmetric in $i,j$ and traceless. Using the identity
\bdm
\ba{l}
\ds \pa{T^{0i}x^k-T^{0k}x^i}x^jh_{ij,k}=
T^{0m}x^nx^j\pa{\delta^l_m\delta^k_n-\delta^k_m\delta^l_n}h_{lj,k}\\
\ds \stackrel{d=3}{=}-\epsilon_{imn}x^mT^{0n}x^j\epsilon^{ilk}\times
\frac12 \pa{h_{lj,k}-h_{kj,l}}
\,,
\ea
\edm
and the definition of the magnetic part of the Riemann tensor
${\cal B}_{ij}\stackrel{d=3}{\equiv}\frac 12\epsilon_{imn}R_{0jmn}$, one finds the
magnetic quadrupole multipole term
\be
{\cal S}_{mult-m}\stackrel{d=3}{=}\frac 23\int dt \left(J^{ij}{\cal B}_{ij}+\ldots\right)\,.
\ee
To avoid the use of the Levi-Civita tensor $\epsilon_{ijk}$ which may not be
straightforwardly generalizable to $d\neq 3$, it is useful to write
the magnetic quadrupole term coupled to gravity as
\be
\label{eq:magq_dual}
\ba{l}
\ds\left.\frac 12\pa{\int_VT^{ij}x^k}h_{ij,k}\right|_{magnetic}=
-\frac 13\paq{\int_Vx^j\pa{T^{0i}x^k-T^{0k}x^i}}\dot h_{ij,k}\\
\ds=\frac 13\paq{\int_Vx^j\pa{T^{0i}x^k-T^{0k}x^i}}\times
\frac 12\pa{\dot h_{kj,i}-\dot h_{ij,k}}=-\frac 13 J^{k|ji}R_{0jik}\,,
\ea
\ee
where an integration by parts has been used and in the last passage the
linearized expression for the Riemann in the TT gauge,
$R_{0ijk}\sim \frac 12\pa{\dot h_{ij,k}-\dot h_{ik,j}}$, has been inserted. Note that
since $R_{0ijk}$ has vanishing traces for pure radiation, one is led to
define the magnetic quadrupole as
\be
\label{eq:jquad}
J^{i|jk}\equiv\int_Vx^j\pa{T^{0k}x^i-T^{0i}x^k}
  -\frac{\delta^{kj}}{d-1}\pa{x_mT^{0m}x^i-\X^2T^{0i}}
  +\frac{\delta^{ij}}{d-1}\pa{x_mT^{0m}x^k-\X^2T^{0k}}\,,
\ee
which is antisymmetric in the index pair $i,k$, traceless, and, in $d=3$,
it is the dual of $J^{ij}$ in (\ref{eq:magq}), as per Eq.~(\ref{eq:Jmdual})
with no ``$R$'' indices.

At the leading order in source internal velocity and self-gravity, it is
actually trivial to match the multipoles to binary constituents' parameters, to
obtain
\begin{equation}
  \ba{rcl}
  \ds I^{ij} &=&\ds \pa{\sum_a m_a  x_a^i  x_a^j}_{TF}\,,\\
  \ds I^{ijk} &=&\ds \pa{\sum_a m_a  x_a^i  x_a^j x_a^k}_{TF}\,,\\
  \ds J^{i|jk} &=&\ds \pa{\sum_a m_a ( x_a^i v_a^k - v_a^i x_a^k) x_a^j}_{TF}\,, 
\ea
\end{equation}
where ``$TF$'' stands for \emph{trace-free}.
Evaluating them in the center of mass frame, one obtains
\begin{equation}
{\cal S}_{\textrm{mult}}\supset-\int {\rm d}t\paq{\frac 12I^{ij}R_{0i0j} +\frac1{6} I^{ijk} R_{0i0j,k} + \frac13J^{b|ia} R_{0iab}}\,,
\end{equation}
where in the TT gauge $R_{0i0j}\simeq -\frac 12\ddot h_{ij}$ and, at leading order,
\be\label{eq:match}
\ba{rcl}
I^{ij}&=&\ds\nu M\pa{x^i x^j}_{TF}\,,\\
I^{ijk}&=&\ds\nu\Delta m \pa{x^i x^j x^k}_{TF}\,,\\
J^{b|ia}&=&\ds\nu\Delta m\paq{x^i x^b v^a - \frac{{\bf x}^2}{d-1}\delta^{ib} v^a-\frac{{\bf x}\cdot {\bf v} }{d-1} \delta^{ia}x^b-\pa{a\leftrightarrow b}} \,,
\ea
\ee
where $\Delta m\equiv M \sqrt{1-4\nu}$ and $\nu\equiv m_1m_2/M^2$ is the
symmetric mass ratio.
This expression for the magnetic quadrupole is valid for any $d$;
hence, its expression can be safely plugged into Eq.~(\ref{eq:tailmr}) to obtain
the correct conservative action.

\section{Explicit form of 5PN tail terms and comparison with previous results}
\label{sec:comparison}
We can now plug Eq.~(\ref{eq:match}) into the
expressions for ${\cal S}^{(e,1)}_{tail}$ and ${\cal S}^{(m,0)}_{tail}$ reported in
Sec.~\ref{sec:calculation} and obtain the tail terms as functions of the
binary variables. We focus on these two contributions as they are the ones
whose lowest order contribution is at 5PN, which is where
\cite{Blumlein:2020pyo,Bini:2021gat} have shown disagreement with the extreme-mass
ratio result.
After reducing higher order derivatives by means of the Newtonian equations of
motion, and neglecting logarithms and the imaginary part, which are anyways
completely determined by the dimensional pole, the result takes the following
form:
\be
\label{eq:taile_expl}
\ds{\cal S}^{(e,1)}_{tail}&\simeq&\ds-\frac{G_N^4 M^3}{189}\nu^2\Delta m^2
\int \frac{{\rm d}t}{r^4} \bigg\{ \frac1\varepsilon\left[\frac{6042}5 v^4-2364 v^2 v_r ^2+1170 v_r^4\right.\nn\\
  &&\ds\qquad\left.+\frac{G_NM}r\pa{\frac{1872}5v^2-432 v_r^2}+\frac{288}5\frac{G_N^2M^2}{r^2}\right]+\frac{157887}{175} v^4-\frac{68442}{35} v^2 v_r ^2\\
&&\ds\qquad+\frac{7167}7 v_r^4+\frac{G_NM}r\pa{\frac{124452}{175}v^2-\frac{26316}{35} v_r^2}+\frac{34848}{175}\frac{G_N^2M^2}{r^2} \bigg\} \,,\nn\\
\label{eq:tailm_expl}
\ds{\cal S}^{(m,0)}_{tail}&\simeq&\ds-\frac{16 G_N^4 M^3}{45}\nu^2\Delta m^2\int \frac{{\rm d}t}{r^4} \pag{\frac1\varepsilon\paq{\frac12 v^4 +v^2 v_r ^2-\frac32 v_r^4}
+\frac{21}{40}v^4+\frac{61}{20} v^2 v_r^2 -\frac{143}{40}v_r^4}\,,
\ee

with $v_r\equiv \frac{{\bf r}\cdot {\bf v}}r$.
Both results are in agreement with the self-force determination of the
scattering angle, as it has been checked in \cite{Blumlein:2020pyo} and then
confirmed in \cite{Bini:2021gat}.
However, the magnetic quadrupole result in Eq.~(\ref{eq:tailm_expl}) is different
from what would be obtained by substituting Eq.~(\ref{eq:magq}) into the
analog result for Eq.~(\ref{eq:tailmr}) obtained in \cite{Foffa:2019eeb},
which is expressed in
terms of the three-dimensional form of the magnetic quadrupole $J^{ij}$, instead
of its dual $J^{k|ji}$, whose introduction is subsequent to \cite{Foffa:2019eeb}.

To understand the origin of the discrepancy, let us replace $J^{a|jb}$ with
its dual expression (\ref{eq:Jmdual}) inside Eq.~(\ref{eq:tailmr}) for $r=0$,
to obtain its \emph{dual} version
\be
   {\cal S}^{(m,0)}_{tail-dual}&\equiv&\int{\rm d}t \pa{\dddot{J_{ij}}}  \epsilon_{ilk}T^{jkbc}\delta^{ld} \epsilon_{adc}\pa{\dddot{J_{ab}}}\,,
\ee
where
\bdm
T^{jkbc}\simeq-\frac{16 G_N^2 E}{135}
\pa{\delta^{jb}\delta^{kc}+\delta^{jc}\delta^{kb}}\times
\pa{\frac 1{\tilde{\varepsilon}}-\frac {49}{20}}
\edm
which is the result reported in \cite{Bini:2021gat}, and which, after using the
three-dimensional relation
$\epsilon_{ilk}\epsilon_{adc}\delta^{ld}=\delta_{ia}\delta_{kc}-\delta_{ic}\delta_{ka}$,
gives
  \be
  \label{eq:tail_dual}
   {\cal S}^{(m,0)}_{tail-dual}&=&\ds-\frac{16G_N ^2 E}{135}\int{\rm d}t \pa{\dddot{J_{ij}}}  \pa{d\delta_{ai} \delta_{bj}-\delta_{ab}\delta_{ij}}
 \pa{\dddot{J_{ab}}}\times\pa{\frac 1{\tilde{\varepsilon}}-\frac {49}{20}} \nn\\
&\simeq&-\frac{16 G_N^2 E}{45}\int{\rm d}t  \left(\frac 1{\tilde{\varepsilon}}-\frac {127}{60}\right)\pa{\dddot{J_{ij}}}^2\,,
\ee
which is the result computed in \cite{Foffa:2019eeb}. 

We note, however, that the identification (\ref{eq:Jmdual}) is valid only for
$d=3$; hence, one is allowed to use it only as long as the final result is finite.
The conservative Lagrangian is indeed finite, but the two separate contributions
coming from the far and the near zone, computed, respectively, in \cite{Foffa:2019eeb} and in
\cite{Blumlein:2020pyo}, have $\varepsilon$ poles in
dimensional regularization. In principle, it is possible to use the far
zone expression in terms of the $J_{ij}$ variables if a similar manipulation
is also performed in the near zone diagrams corresponding to the magnetic
quadrupole tail, along the lines discussed in Appendix B of \cite{Foffa:2019yfl}, which treats in detail the 4PN case.
This is, however, a cumbersome procedure and a much simpler alternative, as done
in the present work, is to completely avoid the use of $J_{ij}$, and stick to
the $d$-dimensional form for the magnetic multipoles in the far zone which
leads to Eq.~(\ref{eq:tailmr}), and not use the result in Eq.~(\ref{eq:tail_dual}).

This explains the discrepancy observed in \cite{Blumlein:2020pyo},
where it is observed that Eq.~(\ref{eq:tail_dual}) is incompatible with the
result obtained via scattering angle/extreme mass ratio limit computation
\cite{Bini:2019nra,Bini:2020hmy}, unless the
coefficient $-127/60$ is offset by $-1/6$. In \cite{Bini:2021gat},
building on Tutti Frutti approach results
\cite{Bini:2019nra,Bini:2020hmy}, 
a different offset is invoked because a different expression for $J_{ij}$ with respect to \cite{Blumlein:2020pyo} 
[namely, the inverse of Eq.~(\ref{eq:Jmdual}), instead of Eq.~(\ref{eq:magq})] is assumed.
However, this is just a formal difference, and both \cite{Blumlein:2020pyo}
and \cite{Bini:2021gat} agree on the only expression
compatible with the scattering angle/extreme mass ratio determination is the
one reported in Eq.~(\ref{eq:tailm_expl}) in the Lagrangian formalism.

\section{Conclusion}
\label{sec:conclusion}
We have formally computed all the tail contributions to the conservative
dynamics of a generic binary system.
Such computation makes use of the recently introduced $d$-dimensional
generalization of the magnetic gravitational multipoles and, when applied to
the specific case of compact binaries,
allowed us to reconcile the effective field theory result with the scattering angle one
at next-to-leading order in the symmetric mass ratio $\nu$ at 5PN order.
Note that finite size effects for spinless black holes vanish in the static
  limit \cite{Binnington:2009bb,Damour:2009vw,Kol:2011vg}; hence, they affect the dynamics at higher than 5PN order,
  which is the lowest order admissible by the \emph{effacement principle} \cite{Damour:1982wm}.
The expression of the scattering angle is still incomplete at
next-to-next-to-leading order in $\nu$, but even the present partial knowledge
is enough to show \cite{Bini:2021gat} that such a sector is incompatible with the
effective field theory near+far zone result obtained so far, which involves,
  among other terms on the far zone side, the nonlinear \emph{memory} process
  due to emission, scattering, and absorption of radiation \cite{Blanchet:1992br}.\footnote{While
  the present work was under review, \cite{Blumlein:2021txe} appeared, further investigating
  the far zone processes contributing at next-to-next-to-leading order in the
   mass ratio.}
We leave the investigation of this remaining mismatch to future work.

\section*{Acknowledgements}
The authors wish to thank Rafael Porto, Johannes Bl\"umlein, and Thibault Damour for useful correspondence.
The work of R.S. is partly supported by CNPq by Grant No. 312320/2018-3.
R.S. would like to thank ICTP-SAIFR FAPESP Grant No. 2016/01343-7.
The work of G. L. A. is financed in part by the Coordena\c{c}\~{a}o de Aperfei\c{c}oamento de Pessoal de N\'{i}vel Superior - Brasil (CAPES) - Finance Code 001.
S.F. is supported by the Fonds National Suisse and by the SwissMap NCCR.

\appendix

\section{Useful formulas}
\label{app:intgrs}
To perform the computations in Sec. \ref{sec:calculation} one needs the
master integral
\be
\label{eq:mi}
\int_{\pp\Q}\frac1{{\cal D}_a}&=&\pa{-p_0^2}^{d-2-a}I_a\,,
\ee
where we adopted the notation ${\cal D}_a\equiv \pa{\pp^2-p_0^2}\pa{\Q^2-p_0^2}\pa{\Q^2-\pp^2}^a$, and
\be
I_a\equiv\frac1{\pa{4 \pi}^d}\frac{\Gamma\pa{a+2-d}\Gamma\pa{a+1-d/2}^2\Gamma\pa{d/2-a}}{\Gamma\pa{2a+2-d}\Gamma\pa{d/2}}\,.
\ee
Note that we have been cavalier about the Green's function boundary conditions,
i.e., about the displacement off the real axis of the Green's functions poles.
We are allowed to do so because, as derived in \cite{Foffa:2021pkg}, casual
or time symmetric boundary conditions give the same result for the real part
of diagrams with up to two Green's functions for radiative modes, and the
imaginary part of the diagrams do not contribute to the conservative dynamics we
are computing here.

In addition one needs the tensor integral
\be
  \label{eq:Strick}
  \ba{l}
  \ds\int_\Q\frac{q_{i_1}\dots q_{i_n}}{\paq{\pa{\K-\Q}^2}^a\paq{\Q^2-(\omega+i\ta)^2}^b}=
  \sum_{m=0}^{[n/2]}S_{a,b}[n,m]\{[\delta]^m[k]^{n-2m}\}_{i_1\cdots i_n}\,,\\
  \ds S_{a,b}[n,m]\equiv
  \frac{\paq{-\pa{\omega+i\ta}^2}^{d/2-a-b+m}}{2^m\pa{4\pi}^{d/2}}
  \frac{\Gamma(a+b-d/2-m)\Gamma(a+n-2m)\Gamma(d+2m-2a-b)}{\Gamma(a)\Gamma(b)\Gamma(d+n-a-b)}\,.
  \ea
  \ee

  The explicit expression of the Thorne coefficients appearing in
  Eq.~(\ref{eq:flux}) is
\be
\ba{rcl}
\ds f_I(r)&=&\ds\frac{(r+3)(r+4)}{(r+1)(r+2)(r+2)!(2r+5)!!}\,,\\
\ds f_J(r)&=&\ds\frac{4(r+2)(r+4)}{(r+1)(r+3)!(2r+5)!!}\,.
\ea
\ee

\section{Detailed matching procedure for the magnetic moments}
\label{app:Jd}

\begin{figure}
\includegraphics[width=.6\linewidth,angle=-90]{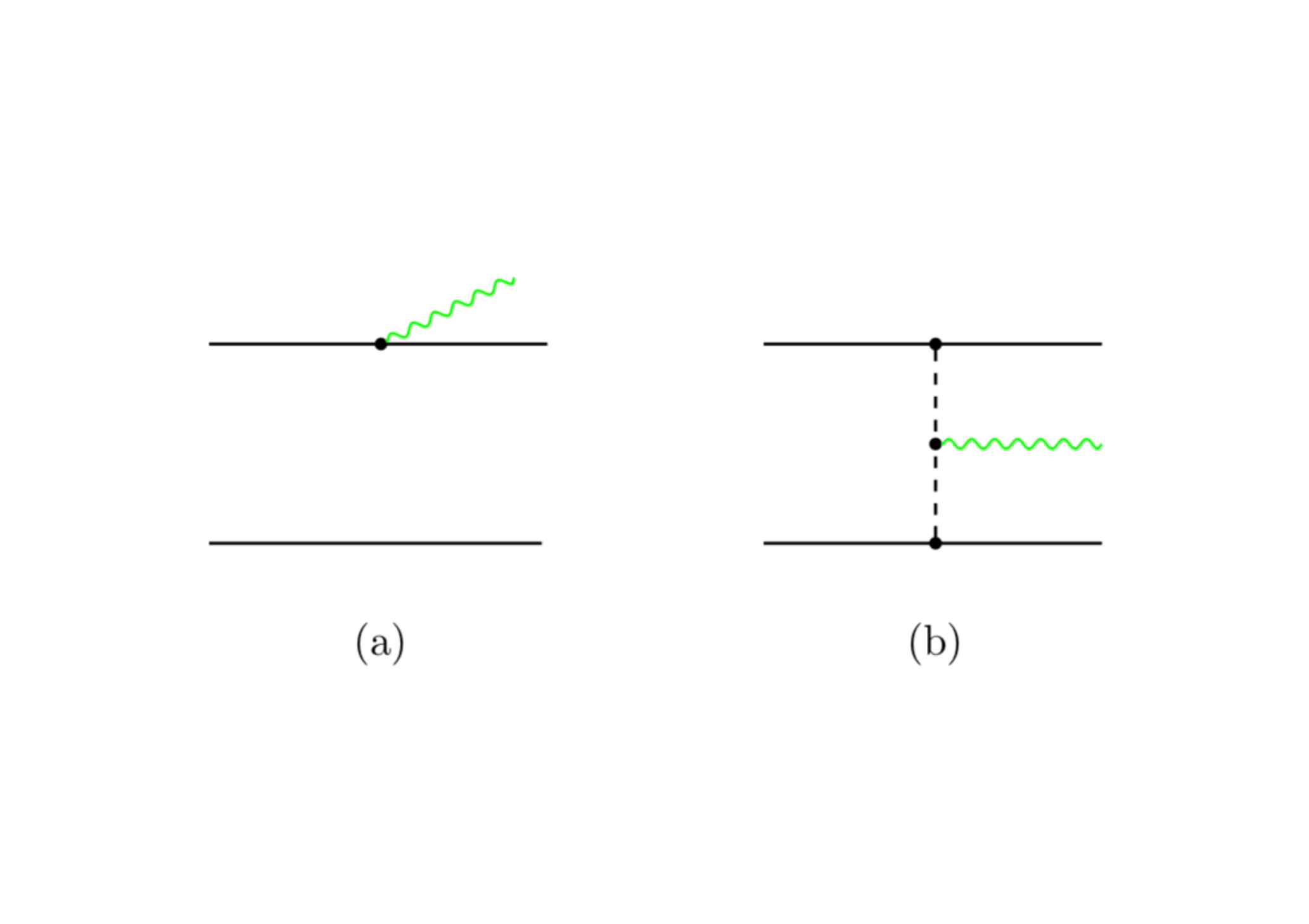}
\caption{Diagrams representing the leading order coupling of a binary system to
  radiative gravitational modes. While diagram (a) yields the first term in Eq. (B1), diagram (b) accounts for the rest of the expression.}
\label{fig:match}
\end{figure}

We give here an alternative, detailed procedure for the matching of the
magnetic moments for the specific case of a binary system, whose coupling
to radiation is described by the diagrams in Fig.~\ref{fig:match}.

We are particularly interested in the coupling of gravitational radiation with
the first moment of $T^{ij}$, which gives rise to the mass octupole and to the
current quadrupole couplings as per Eq.~(\ref{eq:tijxk}).
By expanding up to the first order in the external gravitational mode momentum
$\Q$, and working in the TT gauge,
the sum of the two diagrams gives $1/2 T^{ij}h_{ij}$, with $T^{ij}$ given by
\begin{equation}
T^{ij}(t,\Q)|_{Fig.~\ref{fig:match}} = \sum_a m_a v_a^i v_a^j e^{-{\tt i} \Q \cdot \X_a}
-\frac{(d-2)}{4c_d \Lambda^2} \sum_{a\neq b} \frac{m_a m_b}{r^d} \Theta \left( 1 + \frac{\tt i}{2} \Q\cdot\R_{ab} \right)  r^i r^j  e^{-{\tt i} \Q \cdot \X_a}\,,
\end{equation}
with $\ds\Theta \equiv \frac{\Gamma(d/2-1)}{(4\pi)^{d/2}}2^{d-2}$, and
consequently to
\begin{equation}\label{momTija}
\pa{\int {\rm d}^dx\, T^{ij} x^k}h_{ij,k} = \pa{\sum_a m_a v_a^i v_a^j x_a^k
-\frac{(d-2)}{8c_d \Lambda^2} \sum_{a\neq b} \frac{m_a m_b}{r^d} \Theta ( r_a^k + r_b^k ) r^i r^j   }h_{ij,k} \,.
\end{equation}

Use of the equations of motion
\begin{equation}\label{eomgab}
m_a \ddot{x}_a^i = (-1)^{a+1} (2-d) \frac{m_1 m_2}{2 c_d \Lambda^2}\Theta \frac{r^i}{r^d} \,,
\end{equation}
and some derivation by parts lead finally to the explicit couplings
\begin{equation}
{\cal S}_{\textrm{mult}}\supset\int {\rm d}t \left\{\frac14\left[ \sum_a m_a x_a^i  x_a^j x_a^k \right] \ddot h_{ij,k}-\frac13
\left[ \sum_a m_a ( x_a^i v_a^k - v_a^i x_a^k) x_a^j \right] \dot h_{jk,i}\right\}\,.
\end{equation}

This procedure is fully equivalent for the case of binary systems to the general
textbook approach to the derivation of the multipolar coupling, which has been
sketched in Sec.~\ref{sec:matching} and which we integrate here with some
further details.

The linear coupling of matter $\frac 12T^{\mu\nu} h_{\mu\nu}$ to gravity can be written as
\be
\label{eq:mult_exp}
\ba{rcl}
\ds{\cal S}_{mult}&=&\ds\frac 12\int {\rm d}t\int d^3xT^{\mu\nu}(t,\X)h_{\mu\nu}(t,\X)\simeq\\
&&\ds\frac 12\int {\rm d}t\left\{\pa{\int_VT^{00}}h_{00}+
  \paq{2\pa{\int_VT^{0i}}h_{0i}+\pa{\int_VT^{00}x^i}h_{00,i}}+\right.\\
&& \ds\paq{\pa{\int_VT^{ij}}h_{ij}+\pa{\int_VT^{0i}x^j}\pa{h_{0i,j}+h_{0j,i}}+
    \frac 12\pa{\int_VT^{00}x^ix^j}h_{00,ij}}\\
  &&\left.\ds+\pa{\int_VT^{0i}x^j}\pa{h_{0i,j}-h_{0j,i}}+
  \pa{\int_V T^{ij}x^k}h_{ij,k}+\dots\right\}\,,
  \ea
  \ee
  where all gravitational fields $h$ and their derivatives are understood to be
  evaluated at $\X=0$.
  Working in the center-of-mass frame, the surviving \emph{nonradiative}
  couplings are
  \be
     {\cal S}_{mult-nr}=\frac 12\int dt\pa{ Eh_{00}-J^{i|j} h_{0i,j}}\,,
  \ee
  where we have defined the angular momentum antisymmetric tensor as
  \be
  J^{i|j}\equiv \int_V \pa{x^iT^{0j}-x^jT^{0i}}\,.
  \ee
  We now turn our attention to nonconstant multipoles, and we work for simplicity in the TT gauge.
  By the standard trick of repeated use of the energy-momentum conservation in the
  form $\dot T^{\mu 0}=-T^{\mu i}_{\ \ ,i}$, it is straightforward to derive
  \be
  \label{eq:emt_cons}
  \ba{rcl}
\ds  \int_V T^{ij}&=&\ds\frac 12\frac{{\rm d}^2}{{\rm d}t^2}\pa{\int_V T^{00}x^ix^j}\equiv \frac 12\ddot I^{ij}\,,\\
\ds  \int_V T^{0i}&=&\ds -\frac{\rm d}{{\rm d}t}\pa{\int_V T^{00}x^i}\,.
\ea
\ee
The coupling $\frac 12T^{ij}h_{ij}$, hence, gives rise to the electric quadrupole
coupling $-\frac 12 I^{ij}R_{0i0j}$ term and the next-to-leading-order coupling
$\frac 12 T^{ij}x^k h_{ij,k}$ gives rise to the electric octupole and magnetic
quadrupole couplings as explained in Sec.~\ref{sec:matching}.


\end{document}